\documentclass[prl, twocolumn, showpacs]{revtex4}   %die Option preprint erzeugt doppelten 
\usepackage{epsfig}

\begin{document}

\title{Intrinsic Josephson Effects in the Magnetic Superconductor RuSr$_2$GdCu$_2$O$_8$}

\author{T. Nachtrab}
\author{D. Koelle}
\author{R. Kleiner}
\affiliation{Physikalisches Institut-Experimentalphysik II, Universit\"at T\"ubingen, Auf der Morgenstelle 14, D-72076 T\"ubingen, Germany}

\author{C. Bernhard}
\author{C. T. Lin}
\address{Max-Planck-Institut f\"ur Festk\"orperforschung, Heisenbergstr. 1, D-70569 Stuttgart, Germany}

\date{\today}

\begin{abstract}
We have measured interlayer current transport in small sized RuSr$_2$GdCu$_2$O$_8$ single crystals. We find a clear intrinsic Josephson effect showing that the material acts as a natural superconductor-insulator-ferromagnet-insulator-superconductor superlattice. So far, we detected no unconventional behavior due to the magnetism of the RuO$_2$ layers. 
\end{abstract}


\maketitle

%\narrowtext        % damit der Text auf Spaltenbreite schrumpft

%\section{Introduction}

The discovery of the ruthenocuprate RuSr$_2$GdCu$_2$O$_8$ (Ru1212) created enormous interest due to the coexistence of high-$T_c$ superconductivity with $T_c$ values between 15 and 45\,K (depending on preparation conditions) and magnetic order that sets in below $T_{mag} = 125-145$\,K. The crystal structure of the material consists of alternating superconducting cuprate layers and magnetically ordered ruthenate layers \cite{Bauernfeind,Felner,Bernhard,Lynn}.  The magnetic order of the Ru moment is predominantly antiferromagnetic along the $c$-axis with $1.1-1.2\,\mu_B$ \cite{Lynn}. Nevertheless, a sizeable ferromagnetic (in-plane) component of about $0.1-0.3\,\mu_B$ has also been observed, first by magnetization and muon-spin rotation measurements \cite{Bernhard} and subsequently by neutron scattering \cite{Takagiwa} and NMR \cite{Tokunaga} (the latter two experiments were done on isostructural RuSr$_2$YCu$_2$O$_8$ to avoid the complication due to the magnetic Gd ions). Furthermore, it is found that the ferromagnetic component grows rapidly at the expense of the antiferromagnetic one if an external magnetic field is applied. Already at a field of about 2 Tesla the Ru magnetic order is essentially ferromagnetic. This unusual material marks one of the very few examples where the magnetic transition temperature is considerably higher than the superconducting one. Many interesting phenomena may occur in Ru1212, including a spontaneous vortex-state \cite{Bernhard2,Klamut,Williams} or a superconducting order parameter exhibiting $\pi$ phases \cite{Bulaevskii}.

\begin{figure}
\centering
\epsfig{width=7.5cm,file=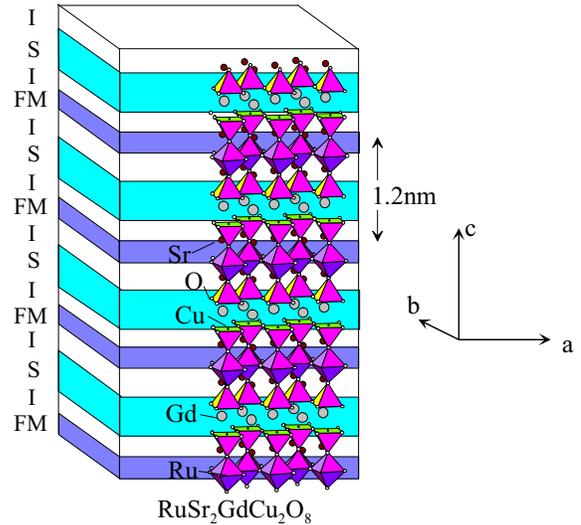}
\caption{Superposition of the crystal structure of Ru1212 and the model of a SIFIS multilayer.}\label{fig01}
\end{figure}

An important question concerns the electronic coupling between the magnetic and the superconducting layers which alternate on a (sub)nanometer scale. Relevant information can be obtained from a detailed investigation of the interlayer transport. The RuO$_2$- and CuO$_2$-layers are separated by SrO layers which are likely to be insulating. The structure thus may form a natural SIFIS multilayer structure, where I stands for insulating layers formed by SrO, F for the magnetic RuO$_2$ layers (with a weak ferromagnetic component) and S for the superconducting CuO$_2$ layers (cf. Fig.~\ref{fig01}).
From this point of view, interlayer supercurrents between adjacent CuO$_2$ layers may be of the Josephson type \cite{Houzet}, similar to the interlayer supercurrent in the cuprates Bi$_2$Sr$_2$CaCu$_2$O$_8$ (Bi2212) or Tl$_2$Ba$_2$Cu$_3$O$_{10}$ \cite{Kleiner}. In the latter compounds this intrinsic Josephson effect is well investigated and has been used for many studies \cite{Yurgens}. Regarding Ru1212, far infrared absorption measurements \cite{Shibata} have revealed a low-lying resonance near 8.5\,cm$^{-1}$ ($\approx 1$\,meV) which has been attributed to longitudinal Josephson-plasma oscillations. To gain more information on the interlayer transport mechanisms, experiments on single crystals are required. Due to the lack of large single crystals, so far most experiments (transport or other) have been performed on polycrystalline samples. In this paper we report on interlayer transport experiments on small-sized Ru1212 single crystals. The measurements show that there is an intrinsic Josephson effect which turns out to be very similar to the intrinsic Josephson effect in cuprates. Within the limited magnetic field and temperature range investigated there is no clear sign of any unconventional behavior not known from cuprate intrinsic Josephson junctions. The fact that intrinsic Josephson effects can be observed, however shows that the crystal quality, although certainly not perfect, is good enough to validate a model of a Josephson coupled SIFIS multilayer. 

%\section{Measurements}

\begin{figure}
\centering
\epsfig{file=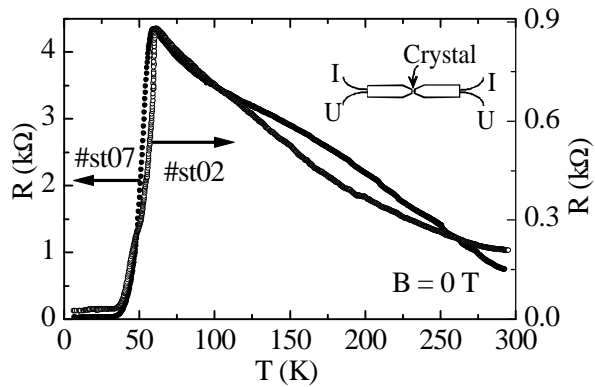}
\caption{Temperature dependence of $c$-axis resistance of two Ru1212 single crystals (open symbols: sample {\it st02}, closed symbols: sample {\it st07}). The inset shows a schematic drawing of the contacting rods.}\label{fig02}
\end{figure}

Ru1212 single crystals were grown by a self flux method as described in \cite{Lin}. The structure and the composition where confirmed by XRD and EDX measurements. Magnetisation measurements on a set of selected single crystals revealed a spontaneous magnetization below $T_{mag}\approx 135$\,K and the onset of a sizeable diamagnetic signal below $T_c \approx 57\,K$ (cf. ref.\,\cite{Lin}). Such a $T_c$ value is somewhat higher than reported for polycrystalline Ru1212 samples and might be explained by slight variations in the Cu, Ru and oxygen content inside the single crystals \cite{Klamut2}. For the transport experiments crystals with an in-plane size between 20\,$\mu$m and 100\,$\mu$m and a thickness between 15\,$\mu$m and 30\,$\mu$m were selected. A 100\,nm thick gold layer was evaporated on both $ab$ faces of the crystal. The crystal was subsequently annealed at 450$^\circ$C for 8\,h to provide a good electrical contact between Au and the crystal surface. Similar to early experiments on cuprate single crystals \cite{Kleiner} the ruthenocuprate crystals were clamped between two contact rods allowing for a two-point interlayer measurement (cf. Fig.~\ref{fig02}, inset). All wires were low pass filtered to minimize environmental noise. Magnetic fields of up to 6\,T could be applied at variable angles relative to the $ab$ planes.
 
In total we successfully measured ten samples so far, all showing similar electrical properties. In this paper we discuss mainly data from three representative crystals, samples {\it st02}, {\it st07} and {\it st21}. The crystals had dimensions $75 \times 75 \times 20\,\mu$m$^3$, $120 \times 120\times 35\,\mu$m$^3$ and  $55 \times 60 \times 25\,\mu$m$^3$, respectively.

%\section{Results}
Fig.~\ref{fig02} shows the temperature dependence of the $c$-axis resistance of two samples. For {\it st07} the midpoint of the superconductive transition is at 51\,K, with a transition width of 10\,K. Sample {\it st02} has a slightly higher $T_c$ of 54\,K. Here, the transition shows a foot-like structure below 50\,K. The residual resistance at low temperatures is due to the contact resistance between the contacting Au layer and the crystal. As can be seen, it amounts to only a few per cent of the total resistance and is thus not a major concern. For both samples $R(T)$ exhibits a maximum near 60\,K. At the resistance maximum we calculate a resistivity $\rho_c$ of 25~$\Omega$\,cm for {\it st02}, and 170~$\Omega$\,cm for {\it st07}. However, these numbers should be taken with some care. The crystals were not perfectly rectangular in shape. Also, we cannot rule out that the contacting Au layer partially shunted the sidewalls of the crystals. We thus have an uncertainity of at least 50\,\% in determining $\rho_c$. Nonetheless one can state that the observed  maximum $\rho_c$ values are of the same order as the ones observed in e. g. Bi2212. 

\begin{figure}
\centering
\epsfig{file=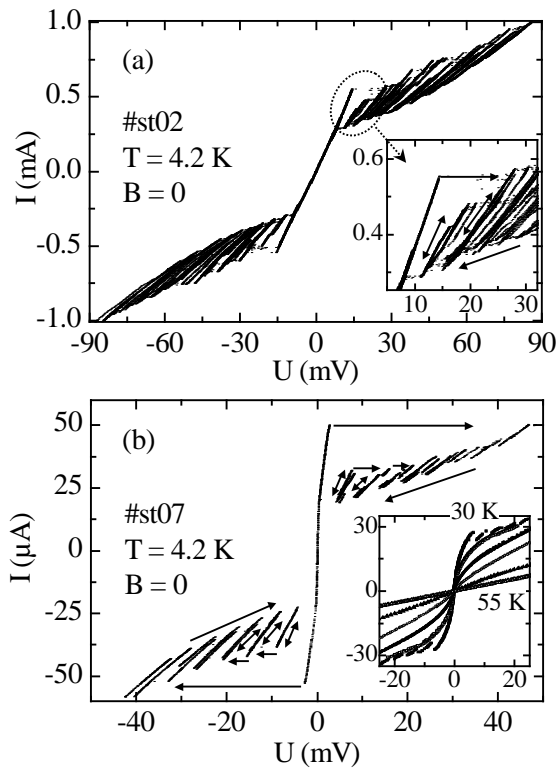}
\caption{Current voltage characteristics of Ru1212 single crystals, with currents applied in $c$ direction. Inset in a) shows an enlargement of the region indicated by the dashed contour. Inset in b) shows $IV$ characteristics at temperatures between 30\,K and 55\,K in steps of 5\,K.}\label{fig03}
\end{figure}

Fig.~\ref{fig03} shows current voltage ($IV$) characteristics of the two samples at 4.2\,K in zero magnetic field. There is the typical multi-branched structure known from the intrinsic Josephson effect in cuprates. For example, each Bi2212 intrinsic Josephson junction exhibits a bistable  $IV$ characteristic. The junctions within a stack can be switched to the resistive state independently of the others; the overall $IV$ characteristic measuring the total voltage across the stack thus consists of many resistive branches differing by the number of junctions in the resistive state \cite{Kleiner,Yurgens}.
 
For the Ru1212 sample {\it st02} (Fig.~\ref{fig03}a), increasing the bias current from zero, one initially observes a finite slope which is due to contact resistance. There are two small voltage jumps at currents of respectively 0.27\,mA and 0.36\,mA, followed by a large jump at 0.55\,mA. At these currents the weakest junctions or groups of junctions in the crystal become resistive. Relating the jump at  0.55\,mA to the full in-plane crystal area of $75 \times 75\,\mu$m$^2$ one obtains a critical current density $j_c$ of about 10\,A/cm$^2$ which is a rather low value. Similar to the $c$-axis resistivity also this number should, however, be taken with caution. The crystal surface is not atomically smooth and thus the "weakest" junctions may be related to a much smaller area. The true $j_c$ value may be significantly higher. With a further increase of current, additional voltage jumps occur. Lowering the current on some resistive branch the voltage decreases continuously down to some return current where the voltage jumps backwards. Multiple current sweeps finally yield the $IV$ characteristic of Fig.~\ref{fig03}a.  The observed distribution of switching currents shows that the crystal is not perfectly homogeneous. However, to be fair, one has to say that similar critical current distributions have been observed for many cuprate crystals exibiting the intrinsic Josephson effect. On average, the voltage jump between adjacent branches is $1-2\,$mV which is an order of magnitude lower than for Bi2212. Such low values may indicate tunneling via electric states localized in the barrier layers.
 
The current voltage ($IV$) characteristic of sample {\it st07} at 4.2\,K, as shown in Fig.~\ref{fig03}b, also exhibits multiple branches. At a current of 50\,$\mu$A a voltage jump of about 50\,mV occurs which corresponds to a larger group of junctions. Inner branches could only be traced out by lowering the bias current from this resistive state below the return currents of some junctions. The switching currents of these "inner" branches  are between 30 and 40\,$\mu$A, i. e. significantly lower than the initial switching current at 50\,$\mu$A.  This feature points to a strong interaction between the different junctions, possibly due to Josephson vortex formation. At higher temperatures both the critical currents and the hysteresis in the $IV$ characteristics of all samples decreased. Above about 35\,K the $IV$ curves were non-hysteretic and noise rounded. The inset in Fig.\,\ref{fig03}b shows examples for sample {\it st07}.
 
\begin{figure}
\centering
\epsfig{file=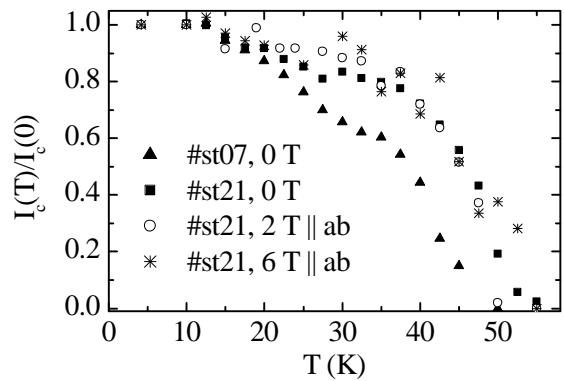}
\caption{Temperature dependence of the critical currents of the first branches of the samples {\it st07} and {\it st21}, respectively. The curves are normalized to their values at 4.2\,K. The midpoint of the zero field superconducting transition is 51\,K  for sample {\it st07} and 55\,K  for sample {\it st21}, respectively.}\label{fig04}
\end{figure}

The temperature dependence of the normalized critical current ($I_c(T)$) of the first branch of samples {\it st07} and {\it st21}, respectively, is shown in Fig.\,\ref{fig04}. For the latter sample measurements were done both in zero magnetic field and in fields of 2 and 6\,T applied parallel to the $ab$ planes. 

Note that, within error bars, in all cases $I_c(T)$ decreases monotonically. There is no indication of a reentrant supercurrent as it has been observed by Ryazanov et al.~\cite{Ryazanov} for metallic superconductor/ferromagnet/superconductor $\pi$-junctions. The data of Figs.~\ref{fig02} to \ref{fig04} thus indicate a "conventional" intrinsic Josephson effect.

We finally discuss the effect of external magnetic fields applied under angles $-90^\circ < \theta < 90^\circ$ relative to the $ab$ planes. At the present stage, the structure of the current voltage characteristics is not regular enough to carefully study minute effects associated with the motion of fluxons. However, an important test can be made on the orientation of the Josephson junctions seen in the $IV$ curves. In a magnetic field there is a voltage contribution due to the motion of Josephson fluxons. When the field is applied at some large angle $\theta$ with respect to parallel orientation, in addition to strings of Josephson fluxons, pancake vortices form penetrating the superconducting layers. In that situation the measured critical currents are small and the voltage is dominated by a large flux flow. When the field is rotated closer to parallel orientation the number of pancakes decreases and the voltage, measured at fixed bias current, becomes smaller (roughly, one can say, that the Josephson fluxons become pinned by the pancakes). Close to parallel orientation, however, there is the lock-in transition \cite{Feinberg} where the $B$ field becomes oriented parallel to the layers.  Now, the Josephson fluxons can move freely and the voltage drop across the layers again increases giving rise to a sharp voltage peak close to parallel orientation. For Bi2212 this effect is routinely used to align the samples with high accuracy with respect to the magnetic field orientation \cite{Hechtfischer}.
\begin{figure}
\centering
\epsfig{file=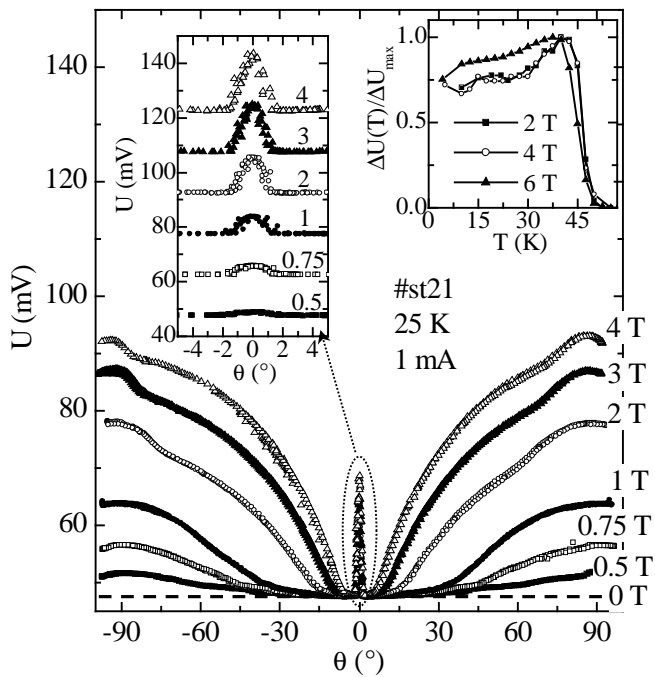}
\caption{Dependence of the voltage across a ruthenocuprate crystal (sample {\it st21}) on the angle of the magnetic field with respect to in-plane orientation at different fixed field amplitudes between 0.5\,T and 4\,T. The sample was biased at 1\,mA. The measurement temperature was 25\,K. Left inset shows an enlargement of the region around parallel orientation (here the curves above 0.5\,T are shifted in steps of 15\,mV). Right inset shows the temperature dependence of the height of the peak at $\theta=0^\circ$, normalized to its maximum value near 35\,K.}\label{fig05}
\end{figure}
Fig.~\ref{fig05} shows corresponding curves for the Ru1212 sample {\it st21}. The characteristic peak near $\theta = 0^\circ$ can be clearly seen (left inset in Fig.\,\ref{fig05}). This peak was obervable at all temperatures up to $T_c$ (right inset in Fig.\,\ref{fig05}). The following can be learned from these curves: First of all, they highlight that fluxon motion takes place in a very similar fashion as in Bi2212. Second, as in Bi2212, the effect is present for fields in the Tesla range showing that there is still a Josephson effect even at such high magnetic fields. Third, the existence of a pronounced peak due to Josephson vortex motion for all temperatures rules out the suppression and reentrance of Josephson coupling due to $0-\pi$ transitions. Finally, by comparing the peak position with the geometric orientation of the crystal, we see that the barrier layers of all Josephson junctions appearing in the experiment are truly oriented in the $ab$-direction of the crystals. Had we measured para\-sitic junctions formed by some cracks or some crystal inhomogeneities in the crystal there would be no reason for a peak to appear near parallel field orientation. We thus can state that we clearly see intrinsic Josephson junctions formed inside the Ru1212 crystals.

%\section{Conclusion}

In summary, we have measured the interlayer current transport in micrometer-sized RuSr$_2$GdCu$_2$O$_8$ single crystals. We find a clear intrinsic Josephson effect, however, for fields up to 6\,T, with no signs of any unconventional behavior due to the magnetic RuO$_2$ layers. This does, however, not mean that such effects do not exist. For example, a spontaneous vortex-state should occur close to $T_c$ where our measurements so far have been strongly affected by thermal fluctuations. Further investigations will show whether or not unconventional phenomena expected for SIFIS multilayers can be observed in this material.

%\section{Acknowledgements}
We gratefully acknowledge financial support by the Deutsche Forschungsgemeinschaft and the Landesforschungsschwerpunktsprogramm Baden-W\"urttemberg.

% now the references. delete or change fake bibitem. delete next three
%   lines and directly read in your .bbl file if you use bibtex.



\end{document}